\documentclass{aastex61}

\shorttitle{Scale invariant jets}
\shortauthors{Liodakis et al.}

\begin{document}

\title{Scale invariant jets: from blazars to microquasars}

\title{Scale invariant jets: from blazars to microquasars}

\correspondingauthor{I. Liodakis}
\email{ilioda@stanford.edu}

\author{Ioannis Liodakis}
\affil{KIPAC, Stanford University, 452 Lomita Mall, Stanford, CA 94305, USA}
\affil{Department of Physics and
Institute for Theoretical and Computational Physics,
University of Crete, 71003,
Heraklion, Greece}

\author{Vasiliki Pavlidou}
\affiliation{Department of Physics and
Institute for Theoretical and Computational Physics,
University of Crete, 71003,
Heraklion, Greece}
\affil{Foundation for Research and Technology - Hellas,
 IESL, Voutes, 7110 Heraklion,
  Greece}

\author{Iossif Papadakis}
\affiliation{Department of Physics and 
Institute for Theoretical and Computational Physics,
University of Crete, 71003,
Heraklion, Greece}
\affil{Foundation for Research and Technology - Hellas,
 IESL, Voutes, 7110 Heraklion,
  Greece}

\author{Emmanouil Angelakis}
\affiliation{Max-Planck-Institut f\"ur Radioastronomie,
 Auf dem H\"ugel 69,
 53121 Bonn, Germany}

\author{Nicola Marchili}
\affiliation{IAPS-INAF,
 Via Fosso del Cavaliere 100, 00133,
 Roma, Italy}

\author{Johann A. Zensus}
\affiliation{Max-Planck-Institut f\"ur Radioastronomie,
 Auf dem H\"ugel 69,
 53121 Bonn, Germany}

\author{Lars Fuhrmann}
\affiliation{ZESS - Center for Sensorsystems,
 University of Siegen, Paul-Bonatz-Str. 9-11,
 57076 Siegen, Germany}

\author{Vassilis Karamanavis}
\affiliation{Max-Planck-Institut f\"ur Radioastronomie,
 Auf dem H\"ugel 69, 53121 Bonn, Germany}

\author{Ioannis Myserlis}
\affiliation{Max-Planck-Institut f\"ur Radioastronomie,
 Auf dem H\"ugel 69, 53121 Bonn, Germany}

\author{Ioannis Nestoras}
\affiliation{Max-Planck-Institut f\"ur Radioastronomie,
 Auf dem H\"ugel 69, 53121 Bonn, Germany}

\author{Efthymios Palaiologou}
\affiliation{Department of Physics and 
Institute for Theoretical and Computational Physics,
University of Crete, 71003,
Heraklion, Greece}

\author{Anthony C. S. Readhead}
\affiliation{Cahill Center for Astronomy and Astrophysics,
 California Institute of Technology, 1200 E California Blvd,
 Pasadena, CA 91125}

\begin{abstract}

Black holes, anywhere in the stellar-mass to supermassive range, are often associated with relativistic jets. Models suggest that jet production may be a universal process common in all black hole systems regardless of their mass. Although in many cases observations support such hypotheses for microquasars and Seyfert galaxies, little is known on whether boosted blazar jets also comply with such universal scaling laws. We use uniquely rich multiwavelength radio light curves from the F-GAMMA program and the most accurate Doppler factors available to date to probe blazar jets in their emission rest frame with unprecedented accuracy. We identify for the first time a strong correlation between the blazar intrinsic broad-band radio luminosity and black hole mass, which extends over $\sim$ 9 orders of magnitude down to microquasars scales. Our results reveal the presence of a universal scaling law that bridges the observing and emission rest frames in beamed sources and allows us to effectively constrain jet models. They consequently provide  an independent method for estimating the Doppler factor, and for predicting expected radio luminosities of boosted jets operating in systems of intermediate or tens-of-solar mass black holes, immediately applicable to cases as those recently observed by LIGO.

\end{abstract}


\section{Introduction}\label{introd}

Blazars constitute unique laboratories to study extreme astrophysics, from relativistic magnetohydrodynamics and shocks, to particle acceleration, ultra-high energy cosmic rays and neutrino production. They are divided into two sub-classes: Flat Spectrum Radio Quasars (FSRQs) and BL Lac objects (BL Lacs) and are most famous for their extreme variability, apparent superluminal motion and $\gamma$-ray loudness since they comprise the largest detected population of the {\it Fermi} $\gamma$-ray observatory \citep{Acero2015}. Understanding the relativistic highly collimated plasma outflows of blazar jets has proven extremely difficult due to the relativistic effects dominating their emission from radio to $\gamma$-rays \citep{Blandford1979}. These relativistic effects are quantified by the Doppler factor  $\delta=[\gamma (1-\beta\cos\theta)]^{-1}$, where $\gamma$ is the Lorentz factor ($\gamma=(\sqrt{1-\beta^2})^{-1}$), $\beta$ is the velocity of the jet in units of speed of light, and $\theta$ is the angle between their jet axis and the observer's line of sight. Even a small spread in the values of $\gamma$ and $\theta$ among blazars results in a large spread in observed properties, severely complicating the search for empirical correlations that can confirm or constrain jet models. If $\delta$ could be confidently estimated, however, these relativistic effects could be corrected for and blazar jets could be studied in their emission rest frame.

Several methods have been proposed for estimating blazar Doppler factors, but they frequently yield discrepant results \citep{Liodakis2017-II}. Given the numerous assumptions entering each method, it is often challenging to identify the most accurate estimate for any given blazar. Two recent breakthroughs have, however, made such a task tractable for the first time. First, through population modeling of unbiased blazar samples \citep{Liodakis2015}, Doppler factor estimates based on variability studies and the assumption of equipartition between synchrotron emitting particles and magnetic field \citep{Readhead1994,Lahteenmaki1999-III,Hovatta2009} were shown to be the most accurate \citep{Liodakis2015-II}. Second, multi-frequency F-GAMMA radio data have recently enabled the calculation of the highest-ever accuracy variability Doppler factors for 58 well-studied blazars \citep{Liodakis2017}.

Correlations between the BH mass ($M_{\rm BH}$) and  monochromatic radio flux density, or the
monochromatic radio flux density and the X-ray flux density or even all the above combined, have long been established (e.g. \citealp{Merloni2003}). The latter suggests the existence of a plane, termed ``fundamental plane of black hole activity'', which extends from X-ray binaries to active galaxies. { These results support the hypothesis of scale invariance, which implies that the jet formation processes are independent of the black hole mass of the system.  Such a hypothesis has been predicted by theoretical models \citep{Heinz2003}.

\cite{Merloni2003} as well as similar attempts to establish a relation connecting BH-powered jets of different $M_{\rm BH}$ \citep{Nagar2002,Falcke2004,Kording2006,Plotkin2012,Saikia2015} have either explicitly avoided blazars focusing on low-luminosity active galactic nuclei (LLAGNs) or have included a handful of blazars. In the latter case they have either ignored the relativistic effects or specifically chosen their sample to include only low-beamed sources. In cases were there was some treatment of the relativistic effects \citep{Falcke2004,Kording2006,Plotkin2012}, the common practice was to use a single value of $\delta$, when in reality Doppler factors are estimated to range between 1 and 45 based on individual source studies \citep{Lahteenmaki1999-III,Hovatta2009,Fan2009,Liodakis2017} and between 1 and 60 based on population studies \citep{Liodakis2015}. }

However, studies focused on blazars, fully and accurately accounting for their relativistic effects cannot be circumvented, specially if the aim  is to study the physics of jets: it is only in such highly beamed sources where we are certain that the observed spectrum is dominated by the jet emission. For this reason, the extension of such scalings to blazars should have a strong impact on unification models of radio-loud active galactic nuclei and our knowledge of BH-powered jets, increasing manyfold our ability to constrain jet models.

There have been attempts to create similar scaling relations in blazars either using luminosity-luminosity correlations \citep{Nemmen2012,Fan2017} or different jet quantities with the properties of the central engine \citep{Wang2004,Hovatta2010,Bower2015}. However, the relativistic effects hamper any attempt to establish a strong correlation between the rest-frame emission of the jet and the $M_{\rm BH}$ in beamed sources.

Even if we account for the relativistic effects properly, blazars are known to show extreme variability across all frequencies. Therefore, the use of single-epoch measurements of flux densities as a proxy of the source luminosity is highly problematic. In addition, many different mechanisms (e.g. hot corona, synchrotron radiation, and synchrotron self-Compton) contribute to the blazar X-ray flux. Disentangling the different contributions has been so far extremely uncertain at best, making the use of the black hole fundamental plane or similar relations unfeasible.

In this work, in order to overcome all such limitations and probe the physics of blazars we propose a new method to explore the connection between jet power and supermassive black hole in beamed sources by using the rest-frame broad-band radio luminosity ($\mathcal{L}_{\rm int}^{\rm Br-B}$) and the $M_{\rm BH}$.

This work is organized as follows: In section \ref{Data_Analysis} we present the data used in this work, and the manner of their analysis. In section \ref{results} we present the correlation analysis and the best-fit relation. In section \ref{discussion} we discuss our findings and in section \ref{sum-conc} we summarize our results and conclusions.

Throughout this work the adopted cosmology is $H_0=71$ ${\rm km \, s^{-1} \, Mpc^{-1}}$, $\Omega_m=0.27$ and $\Omega_\Lambda=1-\Omega_m$ \citep{Komatsu2009}.

\section{Data \& Analysis}\label{Data_Analysis}

Our sample is a sub-set of the sources monitored by the F-GAMMA program\footnote{http://www.mpifr-bonn.mpg.de/div/vlbi/fgamma/fgamma.html} including all sources with both an available Doppler factor and $M_{\rm BH}$ estimates. The F-GAMMA program monitored a total of about 100 blazars detected by {\it Fermi} for eight years with roughly monthly cadence at ten frequencies from 2.64 to 142.33 GHz \citep{Fuhrmann2016}. The Doppler factors are taken from \citep{Liodakis2017}. In \cite{Liodakis2017}, a novel approach is used to model and track the evolution of the flares through multiple frequencies. That method has provided the most accurate variability Doppler factor estimates to date with an on average 16\% error. We use only estimates with quality indicator ``confident'' or ``very confident'' to ensure robust results. The level of confidence is defined by the number of available frequencies and flares used in the estimation of the Doppler factor (see \citealp{Liodakis2017}). The number of sources in our sample is 26 (20 FSRQs, 4 BL Lacs, and 2 radio galaxies). In order to extend our sample towards lower black hole masses, we also included three $\gamma$-ray loud Narrow Line Seyfert 1 galaxies (NLS1s, \citealp{Angelakis2015}) for which there was a Doppler factor computed with the same method as in \cite{Liodakis2017}. In total our sample consists of 29 sources with $M_{\rm BH}$ spanning from $\sim 10^{6.5}$ to $10^{9.5}$ $M_\odot$.

\subsection{Intrinsic broad-band radio luminosity}\label{br.-band-lum.}

We calculated the intrinsic broad-band radio luminosity ($\mathcal{L}_{\rm int}^{\rm Br-B}$) of our sources using data from the F-GAMMA program.  For each frequency we use the maximum likelihood approach described in \cite{Richards2011} to calculate the maximum likelihood mean flux-density taking into account errors in measurements, uneven sampling and source variability. From that we construct the ``mean'' spectrum which we convert to the intrinsic luminosity using the Doppler factor estimates from \cite{Liodakis2017}. Accounting for the relativistic effects in estimating the rest-frame broad-band radio luminosity in blazars is crucial in order to identify the correct correlation (see section \ref{discussion}).

We convert the maximum likelihood mean flux density for each frequency to the intrinsic luminosity using the following equation:
\begin{equation}
S_\nu=\frac{L_\nu \delta^p}{4\pi{d_L^2}}(1+z)^{1+s},
\label{fluxdensity}
\end{equation}
where $S_\nu$ is the flux-density at a given frequency $\nu$, $L_\nu$ the intrinsic luminosity at that frequency, $\delta$ is the Doppler factor, $d_L$ the luminosity distance, $z$ the redshift, $s$ the spectral index which we have defined as $S\propto\nu^s$, and $p$ is equal to $p=2-s$ for the continuous, and $p=3-s$ for the discrete jet case.

By integrating over frequency we calculate the $\mathcal{L}_{\rm int}^{\rm Br-B}$ for each source assuming either a continuous jet ($\mathcal{L}_{\rm int,c}^{Br-B}$) or a discrete jet ($\mathcal{L}_{\rm int,d}^{\rm Br-B}$). We estimate the uncertainty of the $\mathcal{L}_{\rm int}^{\rm Br-B}$ through formal error propagation taking into account errors in the Doppler factor \citep{Liodakis2017}, flux-density, spectral index, and redshift. For the error in the Doppler factor estimates for the three NLS1s we assumed the 16\% average error from \cite{Liodakis2017}. Since our redshift estimates are all spectroscopic we have assumed a common error of $\delta z$=0.01 .

\subsection{BH mass estimation}\label{BH-mass}

In order to estimate the $M_{\rm BH}$ of our sample we used data from \cite{Torrealba2012} and the scaling relations from \cite{Shaw2012} using the line luminosities of the MgII and Hb spectral lines. The scalings were calibrated using the virial mass estimation method. The uncertainty of the $M_{\rm BH}$ for each source was the result of error propagation. We complemented our sample with values (and their error) from the literature with estimates using the same method for consistency (\citealp{Yuan2008,Shaw2012, Zamaninasab2014}, and references therein). When there was no error estimate for a literature value (\citealp{Yuan2008,Zamaninasab2014}, see Table \ref{tab:blazar_sample}), we assumed as error the average uncertainty of the available estimates using the same spectral line. We avoided using methods that  involve assumptions on the radio flux density, beaming, Spectral Energy Distribution (SED) fitting etc., that could potentially create artificial correlations. All the estimates for the $\mathcal{L}_{\rm int}^{\rm Br-B}$ and $M_{\rm BH}$ used in this work are summarized in Table \ref{tab:blazar_sample}.

\section{Correlation and best-fit model}\label{results}

Assuming that the jet is composed of a series of plasma blobs (discrete jet) we tested for a correlation between the $\mathcal{L}_{\rm int,d}^{\rm Br-B}$ and $M_{\rm BH}$. We used the partial correlation test \citep{Akritas1996} in order to obtain the Kendall correlation coefficient ($\tau$) and its significance taking into account the effects of redshift. The partial correlation test yielded a correlation coefficient of $\tau=0.35$ with  $2\times 10^{-3}$ probability of uncorrelated samples, indicating a moderate correlation. If we assume the jet is a continuous stream of plasma (continuous jet, $\mathcal{L}_{\rm int,c}^{\rm Br-B}$), the correlation becomes significantly stronger ($\tau=0.51$ with $5\times10^{-6}$ probability of no correlation). The correlation is stronger in this case, which is to be expected since the continuous jet is, most likely, the most accurate representation of the jet structure,  as has been repeatedly supported by interferometric radio observations (VLBI, e.g. \citealp{Zensus1997}). For this reason, for the remaining of this work we focus on the continuous jet case. Examining different subsamples there was no case were the p-value was $>10^{-3}$ supporting the robustness of the correlation. For FSRQs alone we find $\tau=0.44$ with p-value $6\times10^{-4}$. Including the BL Lacs the test yielded $\tau$=0.43 with p-value $7\times10^{-4}$, and including the radio galaxies $\tau$=0.45 with probability $2\times10^{-4}$. The significant correlation found in all cases strongly suggests that this is a real trend and not an artifact of the large range of $M_{\rm BH}$.

\begin{figure}[t]
\resizebox{\hsize}{!}{\includegraphics[scale=1]{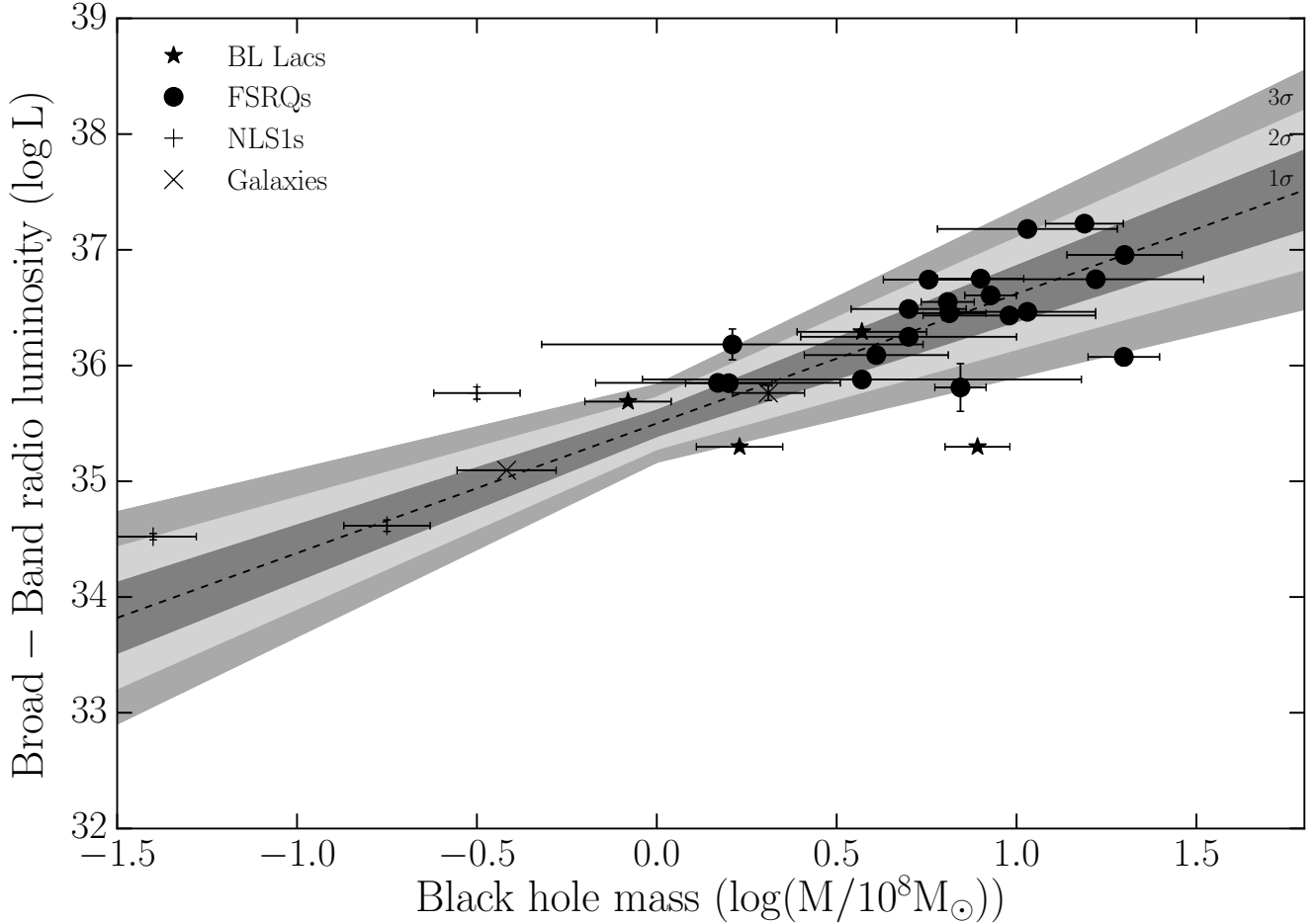} }
 \caption{Intrinsic broad-band radio luminosity ($\mathcal{L}_{\rm int,c}^{\rm Br-B}$) versus $M_{\rm BH}$ (luminosity is in Watts). The dashed line represents the best-fit relation. The grey shaded areas are the 1$\sigma$, 2$\sigma$, and 3$\sigma$ confidence regions respectively, taking into account the error on both slope and intercept.}
 \label{plt_blazar_fit}
 \end{figure}

Having established that a significant correlation exists, we performed a fit between $\mathcal{L}_{\rm int,c}^{\rm Br-B}$ and $M_{\rm BH}$ in log-space. Assuming a linear model of the form $\log\mathcal{L}_{\rm int,c}^{\rm Br-B}=A\times\log(\frac{M}{10^8 M_\odot})+B$. The fit was performed using the BCES bisector methods described in \cite{Akritas1996-II} which takes into account errors in both axis as well as intrinsic scatter. The best-fit line between the $\mathcal{L}_{\rm int,c}^{\rm Br-B}$ and the $M_{\rm BH}$ would then be,

\begin{eqnarray}
\log \left(\frac{\mathcal{L}_{\rm int,c}^{\rm Br-B}}{Watt}\right)&=&(1.12\pm{0.13})\times\log(\frac{M}{10^8M_\odot})\nonumber
+(35.5\pm{0.1}).
\label{eq:continuous}
\end{eqnarray}
The $\log\mathcal{L}_{\rm int,c}^{\rm Br-B}$ versus $\log M_{\rm BH}$ plot is shown in Fig. \ref{plt_blazar_fit} together with the best-fit line. The best-fit results do not depend  on the adopted linear best-fit method. For example, if we consider only intrinsic scatter using the bisector method of \cite{Isobe1990}, the best-fit the slope and intercept become $A=1.05\pm 0.01$ and $B=35.55\pm 0.04$. In fact, for any of the methods described in \cite{Akritas1996-II} and \cite{Isobe1990} the results remain consistent, within the uncertainties.

There is some scatter around the best-fit relation which if intrinsic has been taken into account during the fit. To examine whether this scatter is induced by the errors in the measurements or it is intrinsic, we compare the average distance of each measurement from the best-fit relation to the average distance due to the uncertainty. Using the best-fit line, for each measurement of $M_{\rm BH}$ we estimated the predicted intrinsic radio broad-band luminosity and subtracted it from the observed one to determine the distance in the y-axis. We perform the reverse for the x-axis. We calculate the vertical distance for each observation using $z=\sqrt{x^2+y^2}$ and the scatter is estimated as $\rm Sc=\sqrt{(\sum_{i=1}^{N} z_i^2) /N}$. For the expected distance due to error we use the average error on the y- and x-axis and calculate the vertical distance and scatter in the same manner. We found that the scatter of our sample is $\rm Sc=0.56$ whereas the scatter due to error is $\rm Sc_{error}=0.19$ suggesting that the scatter around the best-fit scaling is intrinsic. 

\begin{figure}[t]
\resizebox{\hsize}{!}{\includegraphics[scale=1]{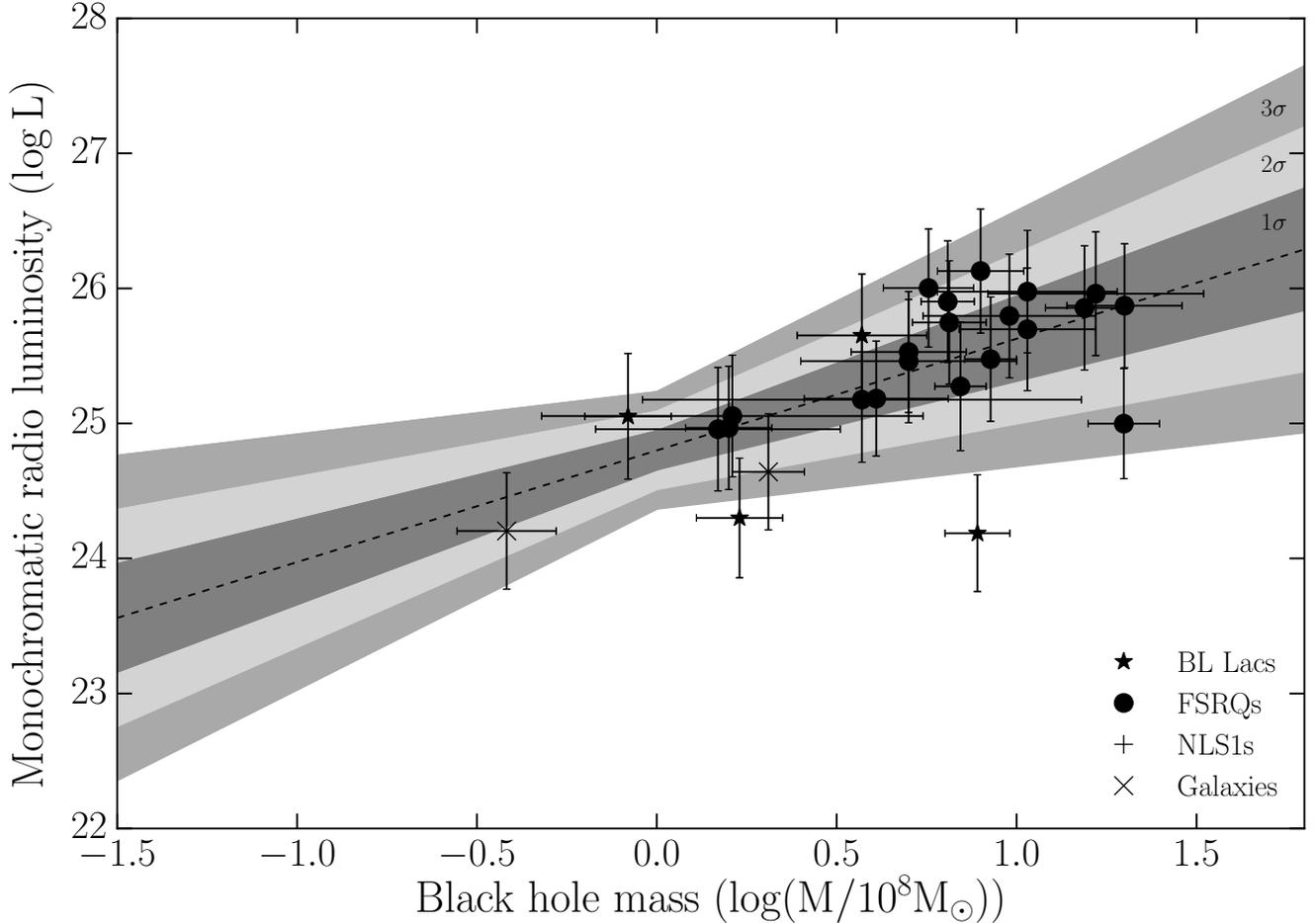} }
 \caption{Intrinsic monochromatic radio luminosity (4.8~GHz) versus $M_{\rm BH}$. The dashed line indicates the best-fit model. Symbols and grey areas as in Fig. \ref{plt_blazar_fit}.}
 \label{plt_monochromatic_fit}
 \end{figure}

Previous studies have opted to use the monochromatic radio luminosity of the jet \citep{Nagar2002,Merloni2003}. We can estimate the mean radio luminosity at a fixed frequency and account for the blazar variability using maximum likelihood. However, observer's frame frequencies are significantly different in the emission rest-frame due to the large redshift span ($z=[0,2]$) of our sample. Since different rest-frame frequencies probe different emission regimes (optically thick or optically thin) in different blazars as well as different regions, the use of single frequency measurements should affect the scatter of the correlation (see also discussion in \citealp{Falcke2004}). Figure \ref{plt_monochromatic_fit} shows the intrinsic monochromatic luminosity at 4.8~GHz versus $M_{\rm BH}$ for our sample.  There is a strong correlation ($\tau=0.44$, p-value $2\times 10^{-4}$) and a slope of $A=0.83\pm 0.17$ and $B=24.8\pm 0.15$. Although we find consistent results between monochromatic and broad-band luminosities, there is a larger scatter ($\rm Sc=0.70$) around the best-fit line in the former case. We therefore conclude that using the broad-band luminosity provides stronger constrains for the  $\mathcal{L}_{\rm int}$-$M_{\rm BH}$ relation.

If jets are indeed scale invariant, then stellar mass BH systems (i.e. microquasars) will have intrinsic broad-band radio luminosities of $\sim 10^{27}-10^{28}$ Watt according to the best-fit relation derived above. To test this prediction, we collected archival data from the literature for five well studied microquasars with available contemporaneous multi-wavelength radio observations. Given the uncertainty in the measurements of different parameters for the microquasars (viewing angle, jet velocity, distance, $M_{\rm BH}$), with each source having multiple estimates in the literature, we used a Monte-Carlo approach to calculate the mean and spread (minimum to maximum) of the $\mathcal{L}_{\rm int,c}^{\rm Br-B}$ and $M_{\rm BH}$ for each source (see Appendix \ref{microquasars}). 

\begin{figure}[t!]
\resizebox{\hsize}{!}{\includegraphics[scale=1]{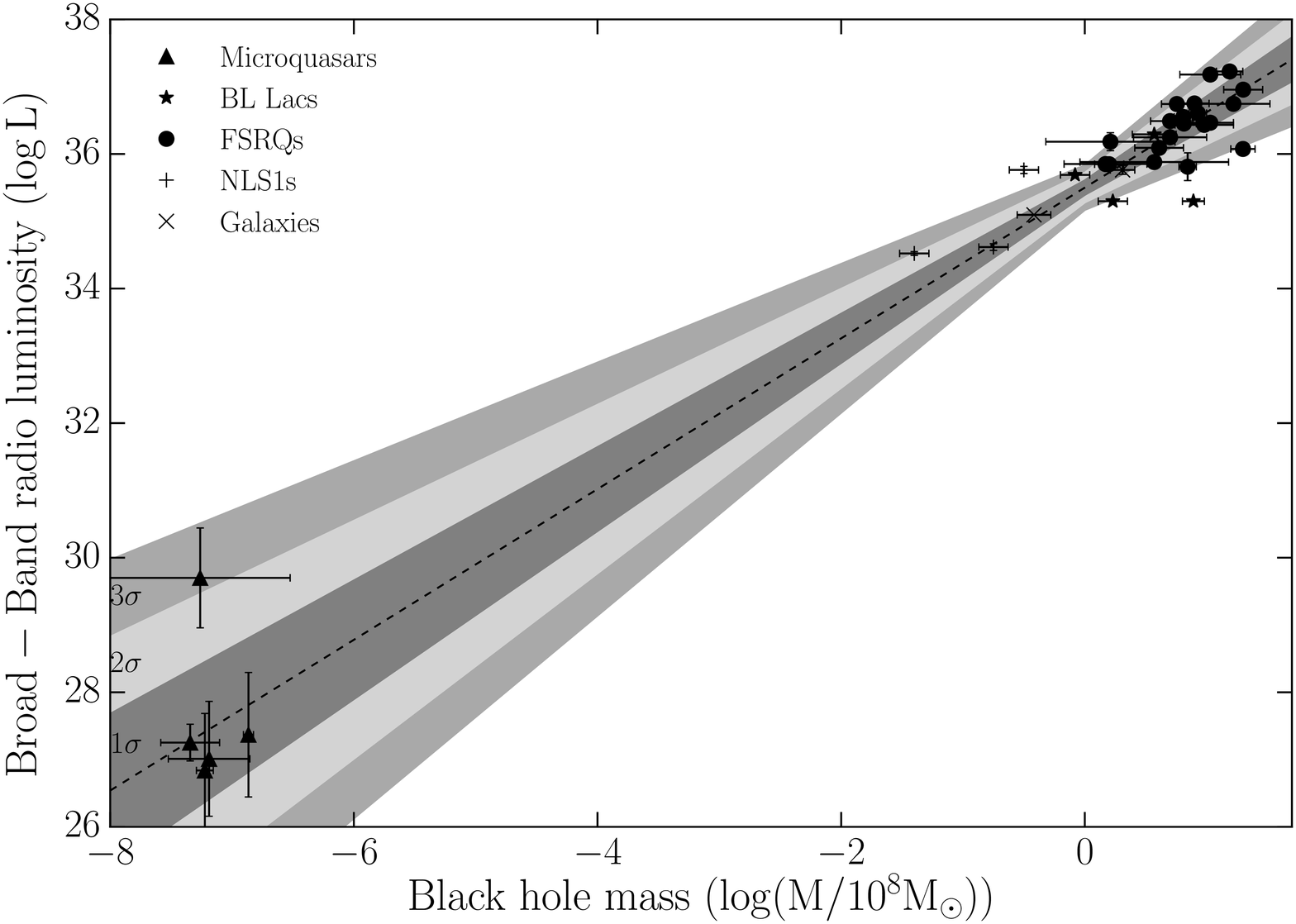} }
 \caption{Intrinsic broad-band radio luminosity ($\mathcal{L}_{\rm int,c}^{\rm Br-B}$) versus $M_{\rm BH}$. Luminosity is in Watts, and $M_{\rm BH}$ in $\rm 10^8M_\odot$. The dashed line represents the best-fit relation using only sources with a supermassive BH. Symbols and grey areas as in Fig. \ref{plt_blazar_fit}. For the microquasars every point represents the mean and the errorbar the spread given the different estimates for each source.}
 \label{plt_scaling}
 \end{figure}

Figure \ref{plt_scaling} shows the position of the microquasars with respect to the best-fit line derived from the supermassive $M_{\rm BH}$ sample. The values represent the mean and the errorbars the spread given the different estimates for each source. All microquasars are consistent with the best-fit $\mathcal{L}_{\rm int,c}^{\rm Br-B}-M_{\rm BH}$ relation for blazars, within the 3$\sigma$ confidence area of the fit, straddling the best-fit line. In fact, four out of the five sources are within 1$\sigma$. While uncertainties of both fit and measurements are quite significant at this mass range, our results suggest that the scaling derived from the blazar sample is in fact a universal scaling extending over at least $\sim$9 orders of magnitude both in $\mathcal{L}_{\rm int,c}^{\rm Br-B}$ and $M_{\rm BH}$.

\section{Discussion}\label{discussion}

\begin{figure}[t]
\resizebox{\hsize}{!}{\includegraphics[scale=1]{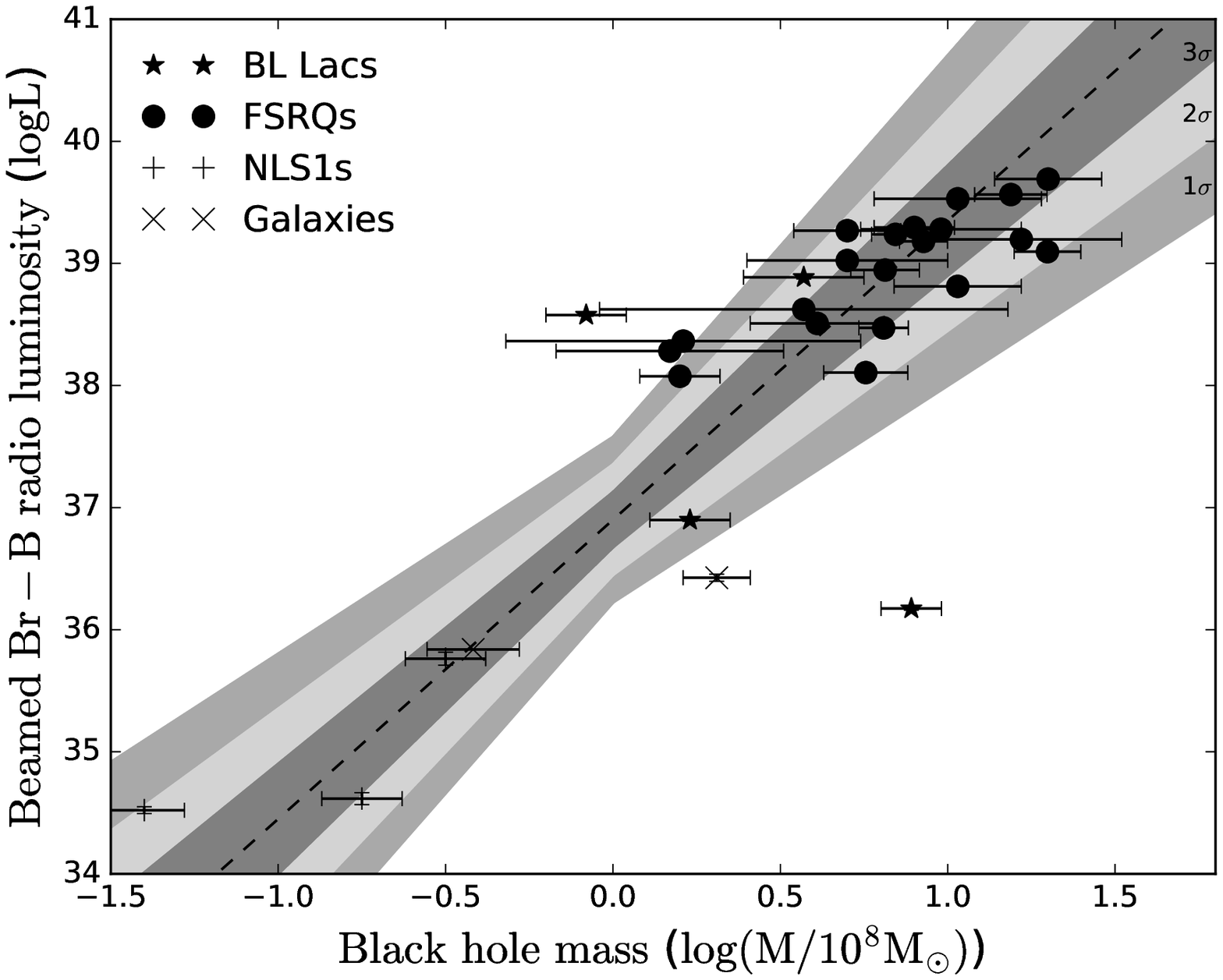} }
 \caption{Observer-frame broad-band radio luminosity versus $M_{\rm BH}$. The dashed line indicates the best-fit model. Symbols and grey areas as in Fig. \ref{plt_blazar_fit}.}
 \label{plt_uncorrected_fit}
 \end{figure}

Taking into account the relativistic effects is essential in order to identify the correct scaling. We investigated the $\mathcal{L}^{\rm Br-B}-M_{\rm BH}$ relation obtained after only correcting for redshift. A correlation is detected, however, the sources appear to occupy unrelated regions of the $\mathcal{L}^{\rm Br-B}-M_{\rm BH}$ plot (Fig. \ref{plt_uncorrected_fit} upper panel). The slope is much steeper (A=2.45$\pm$0.22) with larger scatter. Most importantly the scaling from the supermassive BHs under-predicts the broad-band radio luminosity of microquasars by roughly ten orders of magnitude. Accounting for beaming not only brings all classes with supermassive BHs onto one single line, but it also accurately predicts the position of their stellar mass BH counterparts.

{A non-linear relation between the flux of the jet (and hence the luminosity) and  the $M_{\rm BH}$ ($S_\nu \propto M_{BH}^{1.42}$) was predicted by \cite{Heinz2003}, for typical flat-spectrum core-dominated radio jets and standard accretion scenarios. They also commented that variations in the other source parameters, like the accretion rate (measured in Eddington units), the viscosity parameter and viewing angle will only cause a mass-independent scatter around this relation. This theoretical prediction between radio luminosity and black hole mass is not entirely consistent with the linear relation derived in this work (2.3$\sigma$ difference)}. However, \cite{Heinz2003} pointed that if the jets are powered by spin extraction from the black hole \citep{Blandford1977}, then the jet variables will not only depend on $M_{\rm BH}$ and the accretion rate ({ which was the working assumption in \citealp{Heinz2003}}) but on the spin as well. It is also discussed in \cite{Merloni2003} that the radio luminosity would be sensitive to the spin of the black hole in the case where the process of jet formation is dependent on the spin. Within the framework of the Blandford-Znajek mechanism \citep{Blandford1977}, the luminosity of the jet should be proportional to $M_{\rm BH}$ : $L_{BZ}\propto \dot{m}f(a)^2M_{\rm BH}$ (\cite{Tchekhovskoy2010,Daly2016} and references therein) where $\dot{m}$ is the accretion rate in Eddington units and $f(a)$ is the spin function ($f(a)= a/[1+\sqrt{1-a^2}]$, $a$ being the dimensionless spin parameter). A linear scaling such as the one we find is then expected provided that the product $\dot{m}f(a)^2$ does not vary significant between sources and does not depend on $M_{\rm BH}$.

The intrinsic scatter around the best-fit relation may be due to either $\dot{m}$ or $f(a)$ random variations around the blazar mean $\dot{m}f(a)^2$ product. An additional source of scatter could be variations of the broad-band luminosity. Our estimates of the $\mathcal{L}^{\rm Br-B}$ are limited by the time span of the observations. Although the 8 year dataset of the F-GAMMA program is sufficiently long, it is still possible that exceptional events can occur outside the monitoring period (so that the mean luminosity is actually higher) or alternately a source was unusually active (so that the mean luminosity is actually lower). Cases such as the ones described above are both unlikely and should have a relatively small contribution to the overall scatter.

Our best-fit slope is consistent with those of similar studies on LLAGNs and microquasars ($1.14\pm 0.16$, \citealp{Nagar2002} and $1.23\pm 0.20$, \citealp{Merloni2003}) however with smaller scatter. The larger scatter found in LLANGs could be the result of mild beaming that has not been accounted for (see \citealp{Merloni2003}) or other effects related to the use of single-frequency measurements. Beaming effects can also be responsible for the slightly steeper slope in \cite{Merloni2003}, since the sources whose beaming is important (blazar-like sources) will have systematically higher luminosities compared to the parent population. Thus the slope will be swifted to stepper values. The fact that our results are consistent with those derived for unbeamed sources: (a) provides further support that we are efficiently correcting for the relativistic effects; and (b) provides supportive evidence for the universality of the derived scaling.

The microquasars that agree best with the relation derived from blazars all share the same accretion state while the one at 3$\sigma$ is in the intermediate-soft state (see Appendix \ref{microquasars}). This would suggest that the blazars in our sample are in a similar accretion state as the ones that lie on the best-fit line i.e the hard state. The presence of a big blue bump (BBB) in the SED of FSRQs could indicate a geometrically thin, optically thick disk, contrary to the disk of microquasars in the hard state (e.g.,  \citealp{Done2007}). Although the optical emission from the boosted-jet can dominate over that of the accretion disc and conceal such features in the SED, out of the 20 FSRQs in our sample only 3 sources have a visible BBB. We have verified that excluding these sources from our analysis has no effect on the derived best-fit relation (B remains the same, A=$1.14\pm 0.14$). Same accretion states would also suggest that the Eddington ratios ($R_{Edd}=L_{bol}$/$L_{Edd}$) of the supermassive BH sample would be similar with those of the hard-state microquasars.  Although estimates of $R_{Edd}$ are uncertain for beamed sources, recent studies suggest that $R_{Edd}$ maybe less than 10\% in the majority of blazars and radio galaxies \citep{Ghisellini2010,Chen2015,Xiong2015}. Narrow Line Seyfert 1 are known to have a high $R_{Edd}$, however, recent studies suggest that $R_{Edd}<10$\% in the $\gamma-$ray loud NLS1s as well \citep{Liu2016}. In order to estimate the Eddington ratio for the sources in our sample, we used the \cite{Kaspi2000} relation between Eddington ratio and monochromatic luminosity at 5100 \AA, and we estimated $R_{Edd}$ in nine of them. Although it is not clear whether such relations, which  have been established for radio-quiet objects, are also applicable to radio-loud quasars, we found that eight out of them may have $R_{Edd} < 1$\%, which is almost identical to the Eddington ratios of the microquasars when in their hard state. The only source with $R_{Edd} > 1$\% is one of the three sources with a visible BBB in their SED. Although the lack of $R_{Edd}$ estimates prevents us from drawing strong conclusions, it would appear that, on average, the blazars in our sample share the same accretion state with hard-state microquasars.  It makes sense to compare blazars and microquasars since, after all, they both have jets. We find that $\mathcal{L}^{\rm Br-B}\propto M_{\rm BH}$, for all of them which is consistent with the predictions of Blandford and Znajek mechanism. So, if this is the case, and if the accretion regime is different in hard-state microquasars and some of the blazars in our sample (say powerful FSRQs) then our results suggest that the Blandford and Znajek mechanism can operate, irrespective of the accretion regime (e.g., \citealp{McKinney2005,Tchekhovskoy2010}). As discussed above, differences in $\dot{m}$ (i.e different accretion state) between sources would increase the scatter around the best-fit relation. It will then be interesting to populate our sample with FSRQs with visible and strong BBB, and sources with higher $R_{Edd}$,  in order to examine if the best-fit model moves closer to the microquasar in the intermediate state and the overall scatter decreases. This would suggest that blazars also have two accretion regimes.

The scaling derived in this work provides for the first time an independent method of estimating the Doppler factor starting from first principles. Estimates of the BH mass and the observer's frame broad-band radio luminosity can be used to derive Doppler factor estimates within the scatter of the best-fit relation. These Doppler factor estimates can be used to either reduce the number of free parameters of SED models or, alternately, used to distinguish between acceptable SED models. However, the scaling relation was derived using the mean observer's frame $\mathcal{L}^{\rm Br-B}$ based on the F-GAMMA 8-year dataset. Given the possibility of changes in the Doppler factor during a significant event (bends in the jet, local acceleration e.g. \citealp{Homan2009,Homan2015}), deriving a Doppler factor from single-epoch observations might not yield a representative estimate for that particular event.

\section{Summary \& Conclusions}\label{sum-conc}

Using the multi-wavelength radio light curves of the F-GAMMA program and the most accurate Doppler factors available to date, we identified a strong linear correlation between the intrinsic broad-band radio luminosity of the jets and the black hole mass that extends 9 orders of magnitude to stellar mass black hole systems. Such a universal scaling law is the first ever to bridge observer's and emission rest-frames in beamed sources. 

The scaling derived in this work constitutes an important breakthrough in blazar physics. First, it comprises clear evidence of scale-invariant BH jets and of a connection between the properties of supermassive black holes and the large scale jets they cause in beamed sources. Second, it provides a solid prediction on the $\mathcal{L}_{\rm int,c}^{\rm Br-B}$ of intermediate mass black holes if such exist, and if they form jets. Third, it provides an independent method of estimating the Doppler factor, which will undoubtedly prove an important contribution in constraining SED fitting and the different jet emission models in beamed sources. Fourth, the universality of the scaling would suggest that blazars are in a similar accretion regime as the hard state in microquasars. Finally, our findings point towards the Blandford-Znajek mechanism as the dominant mechanism for jet production in black hole powered jets, and set strong constrains on other potential jet models since they have to reproduce such linear relation.

\begin{acknowledgements}
{The authors would like to thank Andreas Zezas and the anonymous referee for comments and suggestions that helped improve this work.} This research was supported by the ``Aristeia'' Action of the  ``Operational Program Education and Lifelong Learning'' and is co-funded by the European Social Fund (ESF) and Greek National Resources, and by the European Commission Seventh Framework Program (FP7) through grants PCIG10-GA-2011-304001 ``JetPop'' and PIRSES-GA-2012-31578 ``EuroCal''. Our study is based on observations carried out with the 100 m telescope of the MPIfR (Max-Planck-Institut f\"{u}r Radioastronomie) and the IRAM 30 m telescope. IRAM is supported by INSU/CNRS (France), MPG (Germany) and IGN (Spain). I. N., I.M. and V.K. were supported for this research through a stipend from the International Max Planck Research School (IMPRS) for Astronomy and Astrophysics at the Universities of Bonn and Cologne.

\end{acknowledgements}

\appendix

\section{Microquasars}\label{microquasars}

{Our sample of microquasars was selected on the availability of contemporaneous multi-wavelength radio observations. Their data were measured from compact synchrotron jets (hard state) for all sources except Cyg X-3 (the microquasar with the highest luminosity in Fig. \ref{plt_scaling}) whose data were taken from ballistic ejecta during radio outbursts.} Calculating $\mathcal{L}^{\rm Br-B}_{\rm int,c}$ for  microquasars is not trivial, the main obstacle being the lack of robust estimates for key parameters (e.g., distance, Doppler factor), as well as the lack of contemporaneous multi-wavelength radio observations. In most cases the range of available frequencies was confined in the lower frequency range ($<$ 43 GHz). The observed flux density at higher frequencies was the result of extrapolation {by assuming that the spectral index derived from the highest available frequencies of each microquasar remains constant up to 142.33~GHz (which is the highest available frequency for the supermassive BH sample)}.

To estimate the relativistic boosting and revert to the rest-frame, we gathered for each source all available estimates in the literature for the velocity of the jet, distance, and viewing angle. For each source we created a parameter space set by the extrema of the aforementioned quantities. We then uniformly drew a random value for each of these parameters and calculated $\mathcal{L}_{\rm int,c}^{\rm Br-B}$. We repeated this process $10^4$ times and calculated the mean and spread (minimum, maximum) of $\mathcal{L}_{\rm int,c}^{\rm Br-B}$ for each source. The y-values for the microquasars in Fig. \ref{plt_scaling} are the mean, and the y-axis errorbars the spread for each of the sources that resulted from the random sampling.

We follow a similar procedure for the $M_{\rm BH}$, by gathering all the available estimates in the literature and calculating the mean and spread. The x-values for the microquasars in Fig. \ref{plt_scaling} are the mean, and the x-axis errorbars the spread for each of the sources.

The microquasars, parameter ranges, and the references to them are the following:
\begin{table}
\setlength{\tabcolsep}{11pt}
 \centering
 \begin{minipage}{150mm}
 \centering
  \caption{Range of estimates for the different parameters of the microquasars. (1) SIMBAD identification name, (2) alternative source name, (3) Distance in kpc, (4) Lorentz factor ($\Gamma$), (5) Viewing angle ($\theta$), (6) logarithm of the intrinsic broad-band radio luminosity (Watt), (7) logarithm of the BH mass in solar masses.}
  \label{tab:microquasars}
\begin{tabular}{@{}ccccccc@{}}
 \hline
Name &  Alt. Name & Distance & $\Gamma$ & $\theta$ & $\log \mathcal{L }_{\rm int,c}^{\rm Br-B}$  & $\log M_{\rm BH}$ \\
  \hline
V* V1343 Aql & SS 433 & 3-5.5  & 1.02-1.05   & 73-85 & 26.98 - 27.52 & 2.7-7.9 \\
  \hline
V* V1487 Aql & GRS 1915+105 & 8.6-13.7 &  1.01-5.03  & 60-71 & 26.44 - 28.29& 12.4-15\\
  \hline
V* V1521 Cyg & Cyg X-3  & 7.2-12  & 1.05-2.40   & 20-80 & 28.96 - 30.44 & 1-30 \\
 \hline
V* V1033 Sco & GRO J1655-40  & 3-3.5 &  1.04-4.12  & 70-85 &  25.99 - 27.68  & 5.10-7.02 \\
 \hline
V* V404 Cyg &  NOVA Cyg 1989  & 2.25-2.53 & 1.2-10.0   & 46-73 & 26.16 - 27.86 & 3-14\\  
  \hline  
\end{tabular}
\end{minipage}
\end{table}

For SS 433 we used radio data from \cite{Trushkin2003}, and parameter values from \cite{Seward1980,Fender2001,Trushkin2003,Bowler2010,Panferov2010,Cherepashchuk2013} and references therein. 

For GRS 1915+105 we used radio data from \citep{Rodriguez1995}, and parameter values from \cite{Mirabel1994,Fender2001,Fender2003,Fender2004-II,Miller-Jones2005,Punsly2011,Reid2014,Zdziarski2014,Punsly2016} and references therein.

For Cyg X-3 we used radio data from \cite{Koljonen2010}, and parameter values from \cite{Fender1997, Ling2009, Hjalmarsdotter2009, Vilhu2009, Dubus2010, Vilhu2013, Zdziarski2012, Zdziarski2013, Zdziarski2016} and references therein.

For GRO J1655-40 we used radio data from \cite{Hannikainen2000}, and parameter values from \cite{Hannikainen2000, Trushkin2000, Mirabel2002, Fender2003, Fender2004, Narayan2005, Fender2010, Motta2014, Stuchlik2016} and references therein.

For V* V404 Cyg we used radio data from \cite{Gallo2005}, and parameter values from \citep{Cherepashchuk2004,Heinz2004,Miller-Jones2009,Khargharia2010,Xie2014,Siegert2016} and references therein. Since the only available estimate for the jet Lorentz factor \citep{Heinz2004} was a lower limit ($\Gamma > 5$, higher than the alternate estimate) given the typical range of Lorentz factors in these sources  we assumed an upper limit of 10. Assuming a smaller value (e.g., $\Gamma=7$) would result in a difference of maximum values lower than a factor two.

\section{Sample}

\begin{table}
\setlength{\tabcolsep}{10pt}
\tabcolsep=0.15cm
\centering
\begin{minipage}{160mm} 
\centering
 \caption{Blazar sample. (1) F-GAMMA identification name, (2) alternative source name, (3) class (Q is for FSRQs, B for BL Lacs, G for radio galaxies, N for Narrow line Seyferts), (4) redshift, (5)Variability Doppler factor ($\delta_{var}$), (6) Uncertainty of the variability Doppler factor ($\sigma_{\delta_{var}}$) (7) logarithm of the intrinsic broad-band radio luminosity (Watt), (8) uncertainty of the luminosity (9) logarithm of the BH mass in solar masses, (10) uncertainty of the BH mass.}
  \label{tab:blazar_sample}
\begin{tabular}{@{}crcccccccc@{}}
 \hline
Name & ALT-Name & Class & z  &$\delta_{var}$& $\sigma_{\delta_{var}}$ & $\log \mathcal{L }_{\rm int,c}^{\rm Br-B}$ & $\sigma_{\log \mathcal{L } }$ & $\log M_{\rm BH}$ &  $\sigma_{M }$\\
J0102+5824 & 0059+5808 & Q & 0.644  & 21.9  & 3.6   & 35.88 & 0.05 & 8.57$^1$ & 0.61 \\ 
J0136+4751 & 0133+476 & Q & 0.859   & 13.7  & 3.7   & 36.45 & 0.03 & 8.81 & 0.10 \\  
J0237+2848 & 0234+285 & Q & 1.206   & 12.2  & 4.3   & 36.75 & 0.02 & 9.22$^1$ & 0.30 \\ 
J0324+3410 & 1H0323+342 & N & 0.063   & 3.91  & 0.6   & 34.61 & 0.05 & 7.25$^3$ & 0.13 \\
J0418+3801 & 3C111 & G & 0.049   & 2.0  & 0.4   & 35.76 & 0.02 & 8.31 & 0.10 \\ 
J0423-0120 & 0420-014 & Q & 0.916   & 43.9  & 9.2   & 35.81 & 0.21 & 8.84 & 0.07 \\ 
J0433+0521 & 3C120 & G & 0.033   & 2.1  & 0.1   & 35.10 & 0.02 & 7.58 & 0.14 \\ 
J0530+1331 & PKS0528+134 & Q   & 2.070  &  12.9  & 2.5 & 37.18 & 0.05 & 9.03$^2$ & 0.25 \\
J0654+4514 & S40650+453 & Q   & 0.928  &  13.8  & 2.6  & 35.85 & 0.04 & 8.17$^1$ & 0.34 \\ 
J0948+0022 & PMN J0948+0022 & N & 0.583   & 10.12  & 1.6   & 35.76 & 0.05 & 7.5$^3$ & 0.13 \\
J1130-1449 & 1127-145 & Q & 1.184   & 21.9  & 0.0   & 36.07 & 0.07 & 9.30 & 0.10 \\
J1159+2914 & PKS1156+295 & Q & 0.725   &  12.8 & 0.0   & 36.09 & 0.06 & 8.61$^1$ & 0.20 \\  
J1221+2813 & QSOB1219+285 & B & 0.102   & 2.6  & 0.6   & 35.30 & 0.02 & 8.89 & 0.09 \\ 
J1229+0203 & 3C273 & Q & 0.158   & 3.7  & 1.0   & 36.74 & 0.02 & 8.76 & 0.13 \\ 
J1256-0547 & 3C279 & Q & 0.536   & 16.8  & 2.9   & 36.75 & 0.03 & 8.90 & 0.12 \\ 
J1310+3220 & OP+313 & B & 0.997   & 15.8  & 1.7   & 36.29 & 0.02 & 8.57$^1$ & 0.18 \\ 
J1504+1029 & PKS1502+106 & Q   & 1.839  & 17.3   & 2.7 & 36.43 & 0.07 & 8.98$^1$ & 0.24 \\ 
J1505+0326 & PKS1502+036 & N   & 0.408  & 11.32   & 1.8 & 34.52 & 0.03 & 6.6$^3$ & 0.13 \\
J1512-0905 & PKS1510-089 & Q   & 0.360  & 12.3   & 2.8 & 35.85 & 0.06 & 8.20$^2$ & 0.13 \\ 
J1635+3808 & 4C+38.41 & Q   & 1.814  & 20.3   & 2.8 & 36.96 & 0.06 & 9.30$^1$ & 0.16 \\ 
J1642+3948 & 3C345 & Q & 0.593   & 10.4  & 2.9   & 36.46 & 0.02 & 9.03$^1$ & 0.19 \\ 
J1800+7828 & S51803+78 & B   & 0.680  & 21.2   & 5.0 & 35.69 & 0.03 & 7.92$^2$ & 0.13 \\  
J1848+3219 & TXS1846+322 & Q   & 0.798  & 12.1   & 1.4 & 36.18 & 0.13 & 8.21$^1$ & 0.53 \\ 
J1849+6705 & S41849+670 & Q   & 0.657  &  8.1  & 1.4 & 36.55 & 0.02 & 8.81 & 0.07 \\ 
J2202+4216 & BL Lac & B & 0.069   & 6.1  & 0.8   & 35.30 & 0.03 & 8.23$^2$ & 0.13 \\
J2229-0832 & 2227-088 & Q & 1.560   & 21.0  & 0.6   & 36.49 & 0.05 & 8.70$^1$ & 0.16 \\  
J2232+1143 & CTA102 & Q & 1.037   & 15.1  & 4.8   & 36.61 & 0.04 & 8.93 & 0.07 \\ 
J2253+1608 & 3C454.3 & Q & 0.859   & 17.0  & 3.7   & 37.23 & 0.06 & 9.19 & 0.11 \\ 
J2327+0940 & PKS2325+093 & Q & 1.841   & 17.2  & 2.3   & 36.25 & 0.04 & 8.70$^1$ & 0.30 \\ 
\hline
\multicolumn{8}{l}{$^1$\cite{Shaw2012}, $^2$\cite{Zamaninasab2014}, $^3$\cite{Yuan2008}. }
\end{tabular}
\end{minipage}
\end{table}



\end{document}